\documentclass[12pt,preprint]{aastex}

\def\>{$>$}
\def\<{$<$}

\def\simlt{\lower.5ex\hbox{$\; \buildrel < \over \sim \;$}}
\def\simgt{\lower.5ex\hbox{$\; \buildrel > \over \sim \;$}}
\def\ch2{$\chi^{2}$}

\def\ee{\'{e}}

\def\be{\begin{equation}}
\def\ee{\end{equation}}

\def\bB{{\,\mathbf B}}
\def\bE{{\,\mathbf E}}
\def\bj{{\,\mathbf j}}

\def\jB{j_B}

\def\alph{\alpha}
\def\aleff{\alpha_{\rm eff}}
\def\Lgap{L_{\rm gap}}
\def\rhogap{\rho_{\rm gap}}
\def\rhoGJ{\rho_{\rm GJ}}
\def\rpc{r_{\rm pc}}
\def\Epar{E_\parallel}
\def\Et{E_\parallel}
\def\om{\omega_p}
\def\pmax{p_{\rm max}}
  \def\vrot{v_\perp}
  \def\bvrot{{\mathbf v}_\perp}

\def\lrev{l_{\rm rev}}
\def\LA{L_{\rm A}}
\def\Mback{{\cal M}_{\rm back}}
\def\Mscr{{\cal M}_{\rm scr}}

\def\gamp{\gamma_1}
\def\gams{\gamma_\pm}
\def\dzscr{\delta z_{\rm scr}}
\newbox\grsign \setbox\grsign=\hbox{$>$} \newdimen\grdimen \grdimen=\ht\grsign
\newbox\simlessbox \newbox\simgreatbox \newbox\simpropbox
\setbox\simgreatbox=\hbox{\raise.5ex\hbox{$>$}\llap
     {\lower.5ex\hbox{$\sim$}}}\ht1=\grdimen\dp1=0pt
\setbox\simlessbox=\hbox{\raise.5ex\hbox{$<$}\llap
     {\lower.5ex\hbox{$\sim$}}}\ht2=\grdimen\dp2=0pt
\setbox\simpropbox=\hbox{\raise.5ex\hbox{$\propto$}\llap
     {\lower.5ex\hbox{$\sim$}}}\ht2=\grdimen\dp2=0pt
\def\simgt{\mathrel{\copy\simgreatbox}}
\def\simlt{\mathrel{\copy\simlessbox}}

\begin{document}

\title{Polar-cap accelerator and radio emission from pulsars}

\author{Andrei M. Beloborodov\altaffilmark{1}}
\affil{Physics Department and Columbia Astrophysics Laboratory, 
Columbia University \\ 
538 West 120th Street New York, NY 10027
}
                                                                     
\altaffiltext{1}{
Also at Astro-Space Center of Lebedev Physical
Institute, Profsojuznaja 84/32, Moscow 117810, Russia
}

\begin{abstract}
Electric currents $j$ flow along the open magnetic field lines from the 
polar caps of neutron stars. Activity of a polar cap depends on the 
ratio $\alph=j/c\rhoGJ$, where $\rhoGJ$ is the corotation charge density.
The customary assumption $\alph\approx 1$ is not supported by recent 
simulations of pulsar magnetospheres and we study polar caps with arbitrary 
$\alph$. We argue that no significant activity is generated on field lines 
with $0<\alph<1$. Charges are extracted from the star and flow along such 
field lines with low energies. By contrast, if $\alph>1$ or $\alph<0$, a high 
voltage is generated, leading to unsteady $e^\pm$ discharge on a scale-height 
smaller than the size of the polar cap. The discharge can power observed 
pulsars. Voltage fluctuations in the discharge imply unsteady twisting of the 
open 
flux tube and generation of Alfv\'en waves. These waves are 
ducted along the tube and converted to electromagnetic waves, providing a 
new mechanism for pulsar radiation.
\end{abstract}



\section{Introduction}

Corotation of a plasma magnetosphere is impossible beyond the light 
cylinder of a star, and magnetic field lines that extend beyond this 
cylinder are twisted, $\nabla\times\bB\neq 0$. Thus, currents 
$\bj_B=(c/4\pi)\nabla\times\bB$ are induced in the open magnetic flux tubes 
that connect the star (its ``polar caps'') to the light cylinder 
(Sturrock 1971). These currents are approximately force-free and flow along 
the magnetic field $\bB$. A basic question of pulsar theory is what voltage 
develops along the open tube to maintain these currents. It determines the 
dissipated power and $e^\pm$ creation that feeds the observed activity of 
pulsars.

The customary pulsar model assumes that the electric current $\jB$
extracted from the polar cap nearly matches $c\rhoGJ$, where
$\rhoGJ=-{\bf \Omega}\cdot\bB/2\pi c$ is the corotation charge density
(Goldreich \& Julian 1969). The deviation of current from $c\rhoGJ$ was 
calculated as an eigen value of an electrostatic problem and found to be 
small (Arons \& Scharlemann 1979). This is in conflict with recent global 
models of pulsar magnetospheres, which report $|\jB-c\rhoGJ|\sim\jB$ 
(e.g.  Contopoulos et al. 1999; 
Spitkovsky 2006; Timokhin 2006; see Arons 2008 for a review).
A significant mismatch between $\jB$ and $c\rhoGJ$ can be expected 
on general grounds (Kennel et al. 1979). 
Currents $\jB$ are determined by the magnetic-field twisting
near the light cylinder, while $\rhoGJ$ is a local quantity at the polar
cap that is practically independent of $\jB$.

The open tube is surrounded by the grounded closed magnetosphere\footnote{
The closed magnetosphere with $\jB=0$ is expected to have
$\rho=\rhoGJ$ and $\Epar\approx 0$ (e.g. Arons 1979).
}
and may be thought of as a waveguide, filled with magnetized (1D) plasma.
Compared to usual plasma-filled waveguides, it has two special features:
(1) Current $\jB$ is imposed on the tube. The twisted tube extends 
into the star, which is a good conductor and maintains $j_B$.
  ${\rm sign}(\jB)=\pm 1$ corresponds to $\pm$ 
charge flowing outward along the magnetic field lines.
(2) Vacuum has effective charge density $\rho_0=-\rhoGJ$ as Gauss law 
in the rotating frame reads $\nabla\cdot\bE=4\pi(\rho-\rhoGJ)$.
The key 
 dimensionless 
parameter (which can vary along and across the tube) is 
\be
   \alph=\frac{\jB}{c\rhoGJ}.
\ee
In this Letter, we discuss basic properties of the polar-cap accelerator 
with arbitrary $\alph$. First, we discuss what happens without $e^\pm$ 
creation: \S~\ref{sec:slab} studies how the current is extracted from the 
polar cap of a radius $\rpc$ and flows at small heights $z\ll\rpc$ (this 
region is called ``slab zone'' below), and \S~\ref{sec:tube} discusses how 
the flow extends to the region $z>\rpc$ (``thin-tube zone''). We argue that 
the accelerator is inefficient if $0<\alph<1$.

The value of $\alph$ depends largely on the angle $\chi$ between 
${\bf\Omega}$ and $\bB$ at the polar cap. For aligned dipole rotators 
($\chi=0$), magnetospheric models predict $\alph<1$ everywhere on the polar 
cap and $\alpha<0$ near its edge (see Fig.~4 and 5 in Timokhin 2006). 
For orthogonal rotator ($\chi\approx\pi/2$), 
$|\alph|\gg 1$ throughout most of the polar cap.  Generally, the polar cap 
has three regions where $\alph>1$, $0<\alph<1$, and $\alph<0$. 
We propose that pulsar activity originates in the polar-cap regions 
where $\alph^{-1}<1$ (i.e. $\alph>1$ or $\alph<0$) as a high voltage is 
generated in these regions.

We emphasize that the voltage is generated 
because $\nabla\times\bB\neq 0$, not because $\rhoGJ\neq 0$. 
The accelerator works as well if $\rhoGJ=0$ 
(i.e. if $\bf\Omega\perp\bB$ at the polar cap), which corresponds to 
$\alph\rightarrow \pm\infty$. The accelerator height $h\simlt\rpc$ is 
regulated by unsteady $e^\pm$ discharges (\S~\ref{sec:discharge}). 
\S~\ref{sec:Alfven} describes a mechanism for radio emission from unsteady 
discharges.


\section{Acceleration of a charge-separated flow at $z\ll\rpc$} 
\label{sec:slab}

Suppose the current $\jB$ is carried by charges $e$ extracted from the 
polar cap; $e<0$ if $\jB<0$. We assume that this charge-separated flow is 
freely supplied by the star with initial velocity $v\ll c$.
Let us look for a steady state in the rotating frame of the star.
The electrostatic field $\bE$ is vertical near the surface 
(except at the edge of the polar cap). 
Since the displacement current $c^{-1}\partial\bE/\partial t=0$,
the conduction current $\bj$ satisfies
$\bj=\bj_B\equiv(c/4\pi)\,\nabla\times\bB$.
All particles are in the ground Landau state and flow along the magnetic 
field lines. Their motion is governed by the electric-field component 
$\Epar$ (parallel to $\bB$), 
\be
\label{eq:dyn}
   \frac{dp}{dt}=\frac{e\Epar}{mc},
\ee
where $m$ is the charge mass and $p$ is its momentum in units of 
$mc$.  Gravity is negligible here (in contrast to a similar problem for 
closed field lines, Beloborodov \& Thompson 2007).

Let $\psi$ be the angle between $\bB$ and the vertical $z$-axis.
Using $\nabla\cdot\bE\approx dE_z/dz$ for 
the electrostatic field near the surface ($z\ll\rpc$), one finds
from the Gauss law, $d\Epar/dz\approx 4\pi\cos\psi(\rho-\rhoGJ)$.
The flow has charge density $\rho=j/v$ where $v=cp(1+p^2)^{-1/2}$. 
In a steady state, one can rewrite Gauss equation using the convective 
derivative $d/dt\equiv \cos\psi\,v\,d/dz$,
\be
\label{eq:Gauss}
  \frac{d\Epar}{dt}=4\pi j\,\cos^2\psi
                    \left(1-\frac{p}{\alph\sqrt{1+p^2}}\right),
\ee
where $t=0$ corresponds to $z=0$; $\Epar(0)\approx 0$ on the conducting 
surface.
The parameter $\alph(z)$ is not constant, because the magnetic field lines 
are curved (Scharlemann, Arons, \& Fawley 1978) and $\bB$ is changed by 
the frame dragging effect (Muslimov \& Tsygan 1992). At small $z$ one can 
use the linear expansion $\alph(z)\approx\alph_0+\zeta\,z/R$,
where $R$ is the star's radius. We will illustrate with $\zeta=-1$.
Equations~(\ref{eq:dyn}) and (\ref{eq:Gauss}),
are solved together with the equation $dz/dt=cp(1+p^2)^{-1/2}$.
The obtained steady-state models $p(z)$ are summarized in Figure~1.

\begin{figure}
\begin{center}
\epsscale{.80}
\plotone{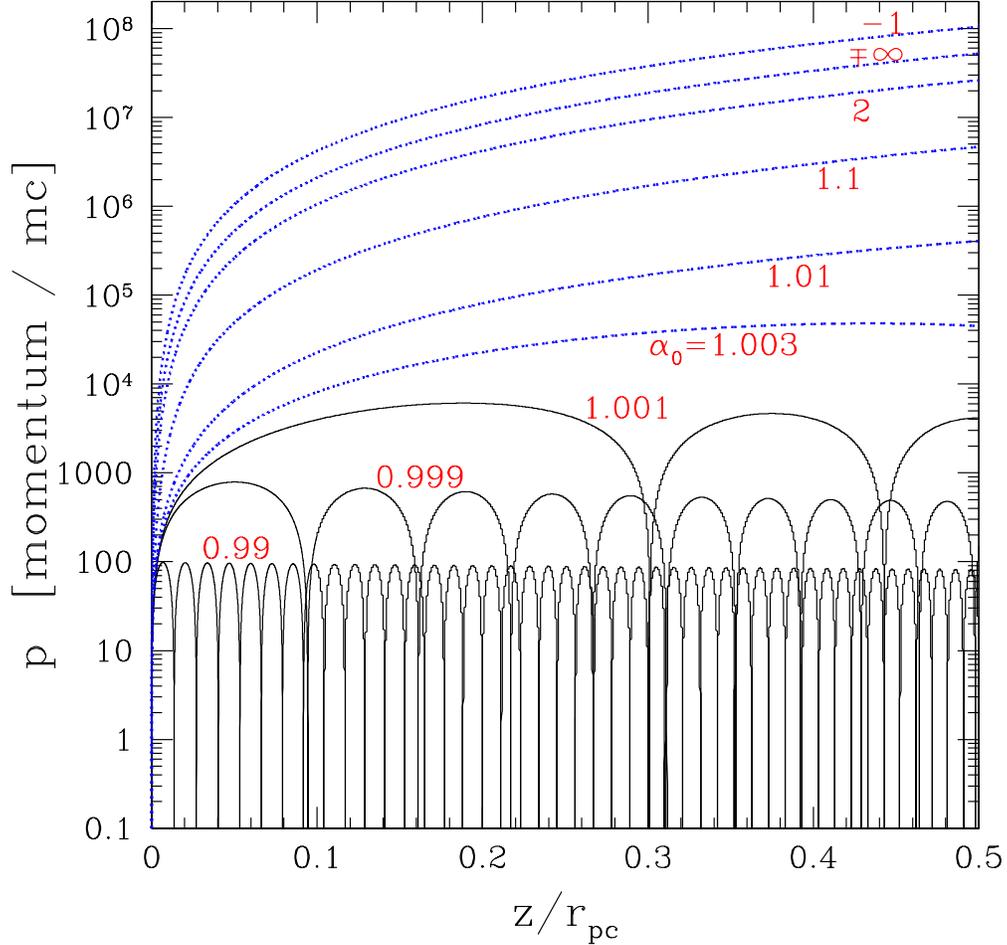}
\label{fig1}
\caption
{Steady-state solutions for the charge flow extracted from the polar cap by 
the self-consistent $\Epar$ in the slab zone $z\ll\rpc$. The magnetic field 
lines are assumed to be perpendicular to the polar cap ($\psi=0$).
Besides $\alph(z)$, the problem has the dimensionless parameter 
$\xi=(4\pi j e/mc^3)^{1/2}\rpc\sim (2\omega_B)^{1/2}\,\Omega\,(R/c)^{3/2}$ 
where $\omega_B=eB/mc$. We assume in the figure $\xi=10^4$ and 
$\rpc/R=10^{-2}$, which approximately corresponds to pulsars with $P\sim 1$~s 
and $B\sim 10^{13}$~G at the polar cap. (In the special case $|\alpha|\ll 1$
one would need to reduce $\xi$ by a factor $\sim|\alpha|$, but we do not 
plot such cases here.) The oscillatory solutions (solid curves) will extend 
to the thin-tube zone $z>\rpc$. The change in
$\alph(z)=\alph_0+\zeta\,z/R$ ($\zeta=-1$ in this example) leads to 
$z_0(z)\neq const$ and $\pmax(z)\neq const$; this effect is visible for 
curves with $|\alpha-1|<|\zeta|\rpc/R\sim 10^{-2}\,|\zeta| P^{-1/2}$. 
The monotonic solutions (dotted curves) illustrate that a high voltage is 
generated if $\alph^{-1}<1$. They do not include $e^\pm$ creation that 
leads to intermittent screening of the accelerator (\S~4). 
Besides, the solutions neglect $\nabla_\perp\cdot\bE_\perp$ in the Gauss 
equation; this term would become comparable to $j/c$ in an unscreened 
accelerator with the maximum $p\sim \xi^2\sim 10^8$.
}
\end{center}
\end{figure}

One can understand the results by considering the simplified model 
with $\alph(z)=const$, which describes a relativistic diode
(e.g. Mestel et al. 1985). Let us also assume $\psi=0$. The first integral 
of equations~(\ref{eq:dyn}) and (\ref{eq:Gauss}) is then given by
\be
\label{eq:fi}
  \frac{\Et^2}{8\pi}=\frac{jmc}{e}
  \left[p-p_0-\frac{1}{\alph}\,
         \left(\sqrt{1+p^2}-\sqrt{1+p_0^2}\right)\right],
\ee
where $p_0\equiv p(0)\ll 1$. $\Epar$ vanishes if $p=p_0$ or 
$p=2\alph/(1-\alph^2)+{\cal O}(p_0)$. The second root is positive 
if $0<\alph<1$; then $p(z)$ oscillates between $p_0$ and 
\be
   \pmax=\frac{2\alph}{1-\alph^2}, \qquad \alpha^{-1}>1.
\ee
The flow starts at $\Epar=0$, with the minimum $p=p_0\ll 1$ and the maximum 
$\rho/\rhoGJ\gg 1$. Maximum $E_{\rm max}^2=8\pi jmce^{-1}(1-\sqrt{1-\alph^2})$ 
is reached where $\rho/\rhoGJ$ passes through 1 ($v=\alpha$).
For $0<\alph\ll 1$ and $p_0\ll\alph$, the solution $p(t)$ 
(of eqs.~\ref{eq:dyn} and \ref{eq:fi}) is a sinusoid 
$p(t)=\alph\left(1-\cos\om t\right)$, where $\om^2\equiv 4\pi e\rhoGJ/m$.
Using $z(t)=\int v\,dt$ one finds $p(z)$. For $0<\alph\ll 1$, $p(z)$ 
is a cycloid with period $z_0=2\pi\alph c/\om$.
For $1-\alph\ll 1$, the period $z_0=2^{3/2}(1-\alph)^{-1}c/\om$. 

By contrast, flows with $\alph^{-1}<1$ (i.e. $\alpha>1$ or $\alpha<0$)
quickly asymptote to
\be
\label{eq:p3}
     p(z)=\left(1-\frac{1}{\alpha}\right)
    \,\left(\xi\,\frac{z}{\rpc}\right)^2, 
 \qquad \xi\equiv\left(\frac{4\pi ej\,\rpc^2}{mc^3}\right)^{1/2}
           \sim 10^4 \left(\frac{B}{10^{12}{\rm ~G}}\right)^{1/2}\,
                     \left(\frac{\Omega}{10{\rm ~Hz}}\right).
\ee
In such flows, $\rho$ does not approach $\rhoGJ$; instead it quickly 
saturates at $j/c$. Then $ed\Epar/dz=4\pi(1-\alph^{-1})ej/c>0$, 
so $e\Epar$ keeps growing $\propto z$, and hence $p$ grows $\propto z^2$. 

The special flow with $\alph=1$ was studied in detail (Michel 1974).
The flow develops $p>1$ at $z>c/\om$, and one finds 
$\Et=(8\pi mc j/e)^{1/2}[1-{\cal O}(1/p)]=const$ and 
$p(z)=\sqrt{2}\; \xi\,z/\rpc$.
This simple solution assumes $\alph(z)=1=const$.
It was modified and extended to $z>\rpc$ by Fawley et al. (1977);
subsequent works studied the effects of slow variations in $\alpha$ near unity.


\section{Extension to the thin-tube zone $z>\rpc$ and time dependence}
\label{sec:tube}

We still discuss here flows with no $e^\pm$ creation. 
The effects of $e^\pm$ are addressed in \S~\ref{sec:discharge}.
Consider a solution from \S~\ref{sec:slab} with $\alph^{-1}>1$ and period 
$z_0\ll\rpc$ (Fig.~1). 
The flow returns to its initial conditions at $z=0$ after passing 
distance $z_0$, i.e., effectively, the location of the polar-cap surface is 
shifted to $z_n=nz_0$ ($n=1,2,...$). Therefore, the flow remains in the slab 
regime and its periodic acceleration/deceleration continues at $z>\rpc$
with a small oscillating potential drop\footnote{
This expression is invalid if $z_0\simgt\rpc$, i.e. if 
$1-\alph\simlt\Delta \equiv 2^{3/2}(c/\om\rpc)
     \sim 6\times 10^{-5}\,P\,B_{13}^{-1/2}(m/m_e)^{1/2}$.
A formal maximum $p=\Delta^{-1}$ (with $\alph$ fine-tuned to 1)
is an overestimate as it neglects the change in $\alph(z)$.
}
$e\Phi/m_ec^2=(1+\pmax^2)^{1/2}-1=2\alph^2(1-\alph^2)^{-1}$.
The parameter $\alph$ and $\Phi$ 
gradually change on scale $\sim$ radius $r\gg z_0$.
Thus, a modest voltage sustains the charge flow along the bundle of field 
lines with $0<\alph<1$. Energy is released with rate $\sim I\Phi$ where 
$I$ is the current along the bundle.
It cannot feed pulsar activity because of small $\Phi$.

A time dependence does not qualitatively change the character of 
the flow with $0<\alpha<1$: $p(z,t)$ remains oscillatory because
$\Epar$ always directs the flow toward $v/c=\alph$. 
Indeed, acceleration to $v/c>\alph$ implies either $\rho/\rhoGJ<1$ or 
$j/\jB>1$. In both cases, $\Epar$ is generated to decelerate the flow:
$\rho/\rhoGJ<1$ implies $e\partial\Epar/\partial z<0$ (Gauss law), and 
$j/\jB>1$ implies $e\partial\Epar/\partial t<0$ (self-induction). 
This negative feedback leads to the oscillatory behavior.
E.g. perturbations with wavevectors 
${\mathbf k}\parallel\bB$ are simple Langmuir oscillations.
Oscillations of amplitude $E_{\rm max}$ can nonlinearly interact with the 
plasma flow, heat it, and lose coherence. Turbulent pulsations of electric 
field can introduce anomalous resistivity and increase voltage. This, however, 
still does not lead to strong particle acceleration needed for pulsar activity.

The basic difference between the regimes $\alpha^{-1}<1$ and $\alpha^{-1}>1$
is due to the interplay of self-induction and electrostatic effects. 
Consider first the case $\alph=0$ (zero current).
Clearly, neutrality $\rho=\rhoGJ$ should be maintained;
the needed charges will be pulled out from the star and a static 
atmosphere can form with $\rho\approx\rhoGJ$.
For a small $\alph>0$, the charges will slowly drift outward and will 
be replenished by new charges extracted from the polar cap by a small $\Epar$. 
This picture can remain qualitatively the same 
with increasing $\alph$ until $\alph$ approaches $1$. When $\alph>1$, the 
extracted charges with density $\rho=\rhoGJ$ are unable to carry the required 
current $\jB$ even if they move with $v=c$; then a large $\Epar$ will be 
induced. In essence, it is the vacuum charge density $\rho_0=-\rhoGJ$ that 
helps avoid high voltages for $0<\alph<1$: then the extracted 
plasma both neutralizes the tube and carries the imposed current. 
Such an accelerator is inefficient because $\Epar$ is dynamically screened 
on the small plasma scale $z_0$. This general argument should hold in full 
3D time-dependent models of the open tube.

Hereafter, we focus on the regime $\alph^{-1}<1$ ($\alph<0$ or $\alph>1$), 
when strong particle acceleration is expected (Fig.~1).
In absence of $e^\pm$ creation, the flow keeps accelerating until it 
approaches $z\sim\rpc$ where $e\Epar$ drops. Consider a simple case 
where $\alph=const$ across the polar cap. If one imagines a global steady 
state with $j=\jB$, the electrostatic potential at $z>\rpc$ 
would be (e.g. Ruderman \& Sutherland 1975), 
\be
  \Phi=S_\perp(\rho-\rhoGJ)=\left(1-\frac{1}{\alph}\right)S_\perp \frac{\jB}{c},
\ee
where $S_\perp$ is the cross section of the open tube. 
$\alph^{-1}<1$ implies $e\Phi>0$, and the accelerated flow 
cannot overcome the potential barrier at the entrance to the thin-tube zone.
It would have to decelerate and reverse at $z\sim\rpc$
(in an attempted steady state, the deceleration would increase density,
making the potential barrier higher). Therefore, $j$ cannot flow steadily. 
When $j/\jB$ drops below unity, the induction current 
$(4\pi)^{-1}\partial\Epar/\partial t=\jB-j$ is generated, and $e\Epar$ 
grows until $j$ recovers back to $j=\jB$.
Thus, in absence of $e^\pm$ creation, the charge-separated flow 
from the polar cap is intermittently reversed by the electrostatic 
barrier at $z\sim\rpc$ and then pushed through by the self-induction effect. 
The particle momentum $p$ achieved in this unsteady regime is
$\pmax\sim(1-\frac{1}{\alph})\,\xi^2$ (eq.~\ref{eq:p3}).


\section{Polar-cap accelerator with $e^\pm$ discharges}
\label{sec:discharge}

Accelerators with $\alpha<0$ or $\alph>1$ are efficient and can easily 
ignite an $e^\pm$ discharge that screens $\Epar$ at a height $h<\rpc$.
The voltage developed in the unscreened ``gap'' $0<z<h$ is
\be
\label{eq:gap}
   \Phi\sim 4\pi (\rho-\rhoGJ)\,h^2
            = 4\pi \left(1-\frac{1}{\alph}\right)\frac{\jB}{c}\,h^2,
 \qquad h<\rpc.
\ee
The gap with $\alpha>1$ was discussed by Sturrock (1971) and Kennel et al. 
(1979). It is different from the vacuum gap of Ruderman \& Sutherland (1975),
which formed because the star was unable to supply charges to keep 
$\rho=\rhoGJ$. The gap described by equation~(\ref{eq:gap}) forms because 
the imposed current $\jB$ oversupplies charge density ($\rho/\rhoGJ>1$ when 
$\alph>1$) or supplies it with opposite sign ($\rho/\rhoGJ<0$ when $\alph<0$).
The electric field in the gap is much stronger than in 
Arons \& Scharlemann (1979) model. The maximal available potential drop 
$\Phi_{\rm max}\sim B\Omega^2R^3/c^2$ is achieved if $h\sim\rpc$.
In observed pulsars $h<\rpc$.

The $e^\pm$ discharge develops as the accelerated primary particles 
(extracted from the polar cap) emit copious $\gamma$-rays that convert to 
outgoing $e^\pm$. The $e^\pm$ Lorentz factor $\gams\sim 10^2-10^3$ is 
much smaller than the primary $\gamp\sim e\Epar z/mc^2$. Therefore, the 
created $-e$ charges (opposite in sign to $\jB$) are easily reversed by 
$\Epar$ and flow back to the polar cap. The reversal length in the unscreened 
$\Epar$ is $\lrev\sim (\gams/\gamp)\,z$.

The backflow would be avoided only if the rate of $e^\pm$ production at $z=h$ 
jumps from 0 to a high value, launching a steady dense $e^\pm$ outflow that 
screens $\Epar$ on a scale $\dzscr<\lrev$. 
This, however, cannot happen for the following reason. 
The maximum charge density that may be created in $\dzscr$ 
by $e^\pm$ outflow without a backflow is\footnote{ 
To avoid the backflow, the $e^\pm$ outflow must be gently polarized, 
with a typical reduction of $\gamma_{-e}$ by a factor $\simlt 2$;
otherwise the dispersion $\Delta\gamma_{-e}/\gamma_{-e}={\cal O}(1)$
would help form a strong backflow.} 
$\rho_\pm\sim -\gams^{-2}\Mscr\jB/c$, where $\Mscr$ is multiplicity 
(number per primary particle) of $e^\pm$ created in $\dzscr$.
The screening of $\Epar$ requires $-4\pi\dzscr\rho_\pm\sim\Epar$, and 
$\dzscr<\lrev$ leads to condition $\Mscr>\gams\gamp^2\,(\rpc/\xi h)^2$. 
It requires a huge $\Mscr$ and cannot be satisfied.\footnote{
Sharpest pair formation fronts occur in accelerators with $\gamp>10^6$, 
where $e^\pm$ are created by curvature photons (e.g. Arons 1979). 
Other channels of pair production (significant at smaller $\gamp$) create 
$e^\pm$ with a smooth distribution over $z$ and help backflow formation.}
One concludes that a strong backflow is inevitable in pair-producing
accelerators with $h<\rpc$. 

A steady-state assumption for the discharge with a strong backflow leads to 
contradictions. First note that, in a steady state, the reversal of $-e$ 
implies $ed\rho/dz>0$. It steepens the growth of $e\Epar(z)$ at 
$h\simlt z\simlt \rpc$ instead of screening $\Epar$. Thus, all charges $-e$ 
created at $z<\rpc$ reverse in the unscreened $\Epar$, creating a backflow 
of a large multiplicity $\Mback\gg 1$. 
The backflowing charges reach the polar cap 
and contribute a fraction $f$ to the total current, leaving the fraction
$(1-f)$ to be carried by charges extracted from the polar cap. In a steady 
state, the acceleration of extracted charges would be described by the same 
equations of \S~\ref{sec:slab} if one substitutes $j=\jB(1-f)$ and replaces 
$\rhoGJ$ by $\rhoGJ+f\jB/c$ (the effective vacuum $\rho_0$ now includes the 
backflow charge density $-f\jB/c$). The new parameter $\aleff$ is given 
by $\aleff=j/(c\rhoGJ+f\jB)=\alph(1-f)/(1+\alph f)$. The presence of 
backflowing charges ($f\neq 0$) changes the condition for gap formation 
from $\alph^{-1}<1$ to $\aleff^{-1}<1$, which is equivalent to
\be
\label{eq:acc}
   f<f_\star\equiv\frac{1}{2}\,\left(1-\frac{1}{\alph}\right).
\ee
The charge density in the relativistic accelerator is 
$\rhogap\approx f\jB/c-(1-f)\jB/c-\rhoGJ=(1-2f-\alph^{-1})\jB/c$. 
It must satisfy $ed\Epar/dz=4\pi e\rhogap>0$, which explains the
condition~(\ref{eq:acc}). 
One then finds: (1) In a steady accelerator with $\alph>1$,
$f$ cannot exceed $\frac{1}{2}$. It permits a backflow multiplicity 
$\Mback\equiv f/(1-f)<1$. A steady $e^\pm$ discharge with $\alph>1$ is 
inconsistent because it would create $\Mback\gg 1$ that switches off the 
polar-cap accelerator. Similarly, a steady state is excluded for $\alph<-1$.
(2) The accelerator with $-1<\alph<0$ could, in principle, have 
$f\rightarrow 1$ and $\Mback\gg 1$. A steady state is, however, implausible
for another reason: a significant rate of ingoing pair creation can be 
expected when $\Mback\gg 1$. It would lead to a runaway loop of $e^\pm$ 
creation in the accelerator, breaking the steady state.
(3) A special case arises if $f\rightarrow f_\star$, which requires 
$\Mback=\Mback^\star\equiv(\alph-1)/(\alph+1)$.
Note that it cannot happen if $-1<\alph<1$. For other $\alpha$,
$\Mback^\star\sim 1$ (unless $\alpha$ is fine-tuned to $-1$).
The accelerator should not permit a steady state with 
$\Mback=\Mback^\star\sim 1$ because the discharge creates $\Mback\gg 1$
even if $f\rightarrow f_\star$ (details will be given elsewhere).

We conclude that the $e^\pm$ discharge is unsteady, repeating on a timescale 
$\sim h/c$. Following a discharge, $f$ exceeds $f_\star$ and voltage is 
screened. Voltage is re-generated when $e^\pm$ leave the polar-cap region
and $f$ 
drops below $f_\star$. In contrast to steady models, $\Epar$ 
is not described by an electrostatic potential $\Phi$. Its time evolution 
obeys the Maxwell equation, $\partial\Epar/\partial t=4\pi(\jB-j)$, which 
expresses the self-induction effect.
The current will fluctuate around $\jB$ (Levinson et al. 2005); 
in addition, the excitation of Alfv\'en waves implies fluctuations of $\jB$ 
itself (\S~5). 
A time-averaged $\Mback\sim 1$ can be expected in such discharges. 
Then the backflow energy flux is comparable to that of the primary particles.
The resulting strong heating of the polar cap can conflict observations, 
posing a problem (cf. Arons 2008).


\section{Alfv\'en-wave pulsar} 
\label{sec:Alfven}

$\bE\neq 0$ in the open tube implies that the tube rotates relative to the 
polar cap (e.g. Ruderman \& Sutherland 1975). The rotational velocity of 
magnetic field lines is given by
$|\bvrot|/c=|\bE\times\bB|/B^2=E_\perp/B$
where $E\ll B$ 
and $E_\perp$ is the component perpendicular to $\bB$.
$\bvrot$ is the drift velocity of plasma;
it may also be associated with rotation of 
magnetic energy since $c\bE\times\bB/4\pi$ is the Poynting flux. 

Consider now a time-dependent state, where the charge density and 
$E_\perp$ fluctuate on a vertical scale $h$ and timescale $t\sim h/c$. 
Such fluctuations will be accompanied by changes in 
$\vrot$, and the magnetic field lines will be twisted with amplitude
\be
   \frac{\delta B_\phi}{B}
   \sim \frac{\delta\vrot t}{h}\sim \frac{\delta\vrot}{c}
   \sim \frac{\delta E_\perp}{B}.
\ee
The corresponding fluctuation in magnetic energy density is
$\delta(B^2/8\pi)=\delta\bB\cdot\bB/4\pi+(\delta\bB)^2/8\pi\sim
(\delta B_\phi)^2/4\pi$.
The time-dependent twisting generates Alfv\'en waves that
propagate along the open tube with nearly speed of light
and carry luminosity 
\be
  \LA\sim \pi \rpc^2\, c\,\frac{(\delta B_\phi)^2}{4\pi}
     \sim\frac{c}{4}\, \rpc^2\, (\delta E_\perp)^2.
\ee
For a gap of height $h<\rpc$ with charge-density fluctuations $\delta\rho$,
one can estimate $\delta E_\perp\sim 4\pi\delta\rho\,h$. This gives
\be
\label{eq:L_A}
   \LA\sim\frac{c(\delta\rho)^2}{\rhogap\jB}\,\Lgap,
\ee
where $\rhogap=\rho-\rhoGJ\sim\Phi/4\pi h^2$, 
$\Lgap=I\Phi$ is the time-averaged power dissipated in the gap, and
$I\sim S_\perp\jB$.
$\LA$ should not exceed $\Lgap$, since it is the gap accelerator that 
provides the energy for shaking the field lines. A related estimate of 
electro-magnetic dipole radiation from sparks is found in Fawley (1978).

The generated Alfv\'en waves are ducted along the field lines and 
eventually suffer Landau damping (e.g. Lyubarsky 1996).
However, part of $\LA$ will convert to transverse electromagnetic waves 
in the emission region, where the modes do not adiabatically track. 
Jil, Lyubarsky, \& Melikidze (2004) show that such conversion 
generally occurs for waves with frequency below the plasma frequency.
This condition is satisfied for Alfv\'en waves and a significant fraction 
of luminosity $\LA$ is expected to escape as radio emission.

Usually considered mechanisms of pulsar radio emission are based on the
two-stream instability in the $e^\pm$ outflow at large radii. The emerging 
radio spectrum is then determined by the plasma physics and possibly by 
curvature emission from plasma bunches created by the two-stream instability.
By contrast, the emission fed by Alfv\'en waves from the gap will have a 
spectrum controlled by the discharge behavior.
One characteristic wavelength may be associated with the gap thickness $h$,
roughly corresponding to frequency $\sim 10$~MHz. However, fluctuations 
$\delta\rho$ will occur also on scales $\lambda<h$, producing waves of 
higher frequencies. Such fluctuations can create
$\delta E_\perp\sim 4\pi\,\delta\rho\,\lambda$, and their contribution to 
Alfv\'en luminosity is 
\be
\label{eq:spectrum}
   \LA(\lambda)\sim 4\pi^2 c\,\rpc^2 \lambda^2\,(\delta\rho)^2
      \sim \Lgap\,\left(\frac{c\rhogap}{\jB}\right)
           \left(\frac{\lambda}{h}\right)^2
           \left(\frac{\delta\rho}{\rhogap}\right)^2.
\ee
A short scale available in the discharge is the screening scale $\lrev\ll h$ 
(\S~\ref{sec:discharge}). The electric field is screened by sudden formation 
of dense fronts of $-e$ backflow on scale $\lambda\sim\lrev$, producing large 
fluctuations $\delta\rho/\rhoGJ\gg 1$, which can generate waves of frequency 
$c/\lrev\sim $~GHz. The spectrum of fluctuations in the $e^\pm$ discharge 
extends from $\sim c/h$ to $c/\lrev$. Numerical simulations of the discharge 
can help find the spectrum of escaping waves and their beaming.


\acknowledgments
This work was supported by NASA grant NNG-06-G107G.


\end{document}